# Test Model for Text Categorization and Text Summarization

Khushboo Thakkar

Computer Science and Engineering
G. H. Raisoni College of Engineering
Nagpur, India

Urmila Shrawankar

Computer Science and Engineering
G. H. Raisoni College of Engineering
Nagpur, India

*Abstract*—**Text Categorization is the task of automatically sorting a set of documents into categories from a predefined set and Text Summarization is a brief and accurate representation of input text such that the output covers the most important concepts of the source in a condensed manner. Document Summarization is an emerging technique for understanding the main purpose of any kind of documents. This paper presents a model that uses text categorization and text summarization for searching a document based on user query.**

## I. INTRODUCTION

Due to the rapid growth of the World Wide Web, information is much easier to disseminate and acquire than before. Finding useful and favored documents from the huge text repository creates significant challenges for users. Typical approaches to resolve such a problem are to employ information retrieval techniques. Information retrieval relies on the use of keywords to search for the desired information. Nevertheless, the amount of information obtained via information retrieval is still far greater than that a user can handle and manage. This in turn requires the user to analyze the searched results one by one until satisfied information is acquired, which is time-consuming and inefficient. It is therefore essential to develop tools to efficiently assist users in identifying desired documents.

One possible means is to utilize text categorization and text summarization. Text categorization is the task of automatically sorting a set of documents into categories from a predefined set and Text Summarization is a brief and accurate representation of input text such that the output covers the most important concepts of the source in a condensed manner. Applying text summarization to a document after finding the document using text categorization saves users time because the user does not need to read through the complete document, instead reading only the summary gives the user the idea about how much useful the document is.

## II. LITERATURE SURVEY

Reynaldo J. Gil-Garcia, Jose M. Badia-Contelles and Aurora Pons-Porrata, [5] presented a general framework for agglomerative hierarchical clustering based on graphs. Specifying an inter-cluster similarity measure, a sub graph of the β- similarity graph, and a cover routine, different hierarchical agglomerative clustering algorithms can be obtained.

QDC algorithm is described in [4] uses the user's query as part of a reliable measure.

Extended Suffix Tree Clustering (ESTC) is explained in [9]. The paper introduces a new cluster scoring function and a new cluster selection algorithm to overcome the problems with overlapping clusters, which are combined with STC to make a new clustering algorithm ESTC.

A graph-based text summarization algorithm is presented in [7]. It is based on finding the shortest path from the first sentence to the last sentence in a graph representing the original text.

Graph-based ranking algorithm TextRank for text summarization is introduced in [11]. The TextRank algorithm is used in this paper in the fourth stage of the model described.

III.   THE MODEL

The model has five stages:

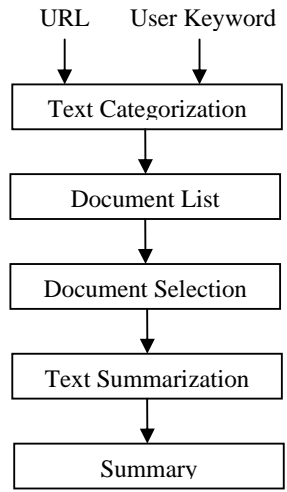

Figure 1.   Five stages of the model

- Text Categorization
- Document List
- Document Selection
- Text Summarization
- Summary

**A.   Text Categorization**

In this stage first the documents are fetched from web according to the URL provided b the user. Then text categorization algorithm is applied to all the documents based on the keyword taken from then user. In this model Query Directed Clustering (QDC) is used for Text Categorization.

**B.   Document List**

After Text Categorization is applied to the documents, we get the list of documents which are related to the keyword which the user has given. This stage displays the list of all those documents.

**C.   Document Selection**

In this stage user is required to select a document which he/she finds useful from the list generated in stage three.

**D.   Text Summarization**

Text Summarization is then applied to the document selected by the user. In this stage TextRank algorithm is used.

**E.   Summary**

The summary generated in the previous stage is displayed for the user.

## IV. QDC ALGORITHM

QDC algorithm is used in the first stage for Text Categorization and it has five phases [4].

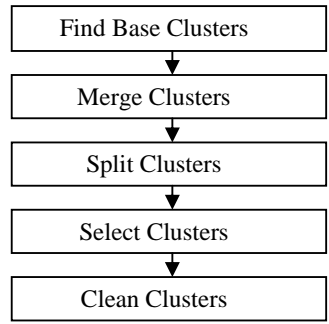

Figure 2. Five phases of QDC

### A. Find Base Clusters

In this stage, first the page is pre-processed that is, all the HTML tags are removed, the stop words are removed and stemming is done. Then the base clusters are formed. QDC constructs a collection of base clusters, one for every word that is in at least 4% of the pages. Then user is asked to enter the keyword for categorization. Then the Normalized Google Distance (NGD) is found between the base clusters and the user keyword. The clusters having large distance are removed.

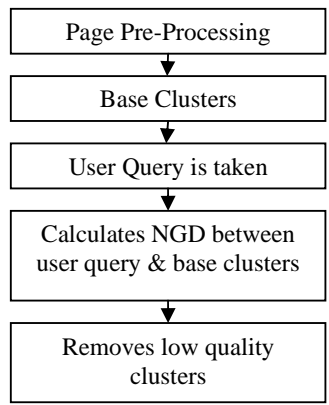

Figure 3. Stages in the first phase of QDC

### B. Merge Clusters

QDC uses Single-link clustering for merging the clusters. Single-link clustering merges together all clusters that are part of the same connected component on the graph.

### C. Split Clusters

Single-link clustering can produce clusters containing multiple ideas and irrelevant base clusters due to cluster chaining. Such clusters need to be split.

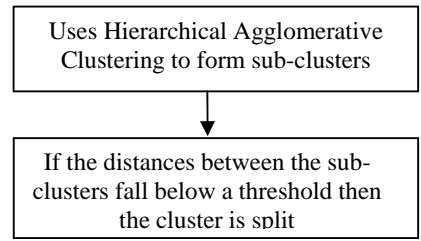

Figure 4.  Stages in the third phase of QDC

This phase uses Hierarchical Aggolomerative Clustering to form the sub-clusters [5]. The algorithm uses a distance measure to build a dendrogram for each cluster starting from the base clusters in the cluster. Each cluster is split by cutting its dendrogram at an appropriate point.

**D.  Select Clusters**

QDC uses Extended Suffix Tree Clustering (ESTC) in the phase [9]. The ESTC cluster selection algorithm uses the heuristic with a 3-step look-ahead hill-climbing search to select a set of clusters to present to the user.

**E.  Clean  Clusters**

This phase is required because clusters can contain pages that cover different topic. QDC computes relevance of each page based on no. & size of cluster's base clusters. Relevance is computed as the sum of sizes of the cluster's base cluster divided by sum of sizes of the entire cluster's base cluster

**V.  TEXTRANK ALGORITHM**

Graph-based ranking algorithms are essentially a way of deciding the importance of a vertex within a graph, based on information drawn from the graph structure. The basic idea implemented by a graph-based ranking model is that of "voting" or "recommendation". When one vertex links to another one, it is basically casting a vote for that other vertex. The higher the number of votes that are cast for a vertex, the higher the importance of the vertex. The score associated with a vertex is determined based on the votes that are cast for it, and the score of the vertices casting these votes.

To enable the application of graph-based ranking algorithms to natural language texts, we have to build a graph that represents the text, and interconnects words or other text entities with meaningful relations. Depending on the application at hand, text units of various sizes and characteristics can be added as vertices in the graph, e.g. words, collocations, entire sentences, or others. Similarly, it is the application that dictates the type of relations that are used to draw connections between any two such vertices, e.g. lexical or semantic relations, contextual overlap, etc.

Regardless of the type and characteristics of the elements added to the graph, the application of graph-based ranking algorithms to natural language texts consists of the following main steps: [11]

- Identify text units that best define the task at hand, and add them as vertices in the graph.

- Identify relations that connect such text units, and use these relations to draw edges between vertices in the graph. Edges can be directed or undirected, weighted or unweighted.

- Iterate the graph-based ranking algorithm until convergence.

- Sort vertices based on their final score. Use the values attached to each vertex for ranking/selection decisions

TextRank does not require deep linguistic knowledge, nor domain or language specific annotated corpora, which makes it highly portable to other domains, genres, or languages.

## VI. RESULTS

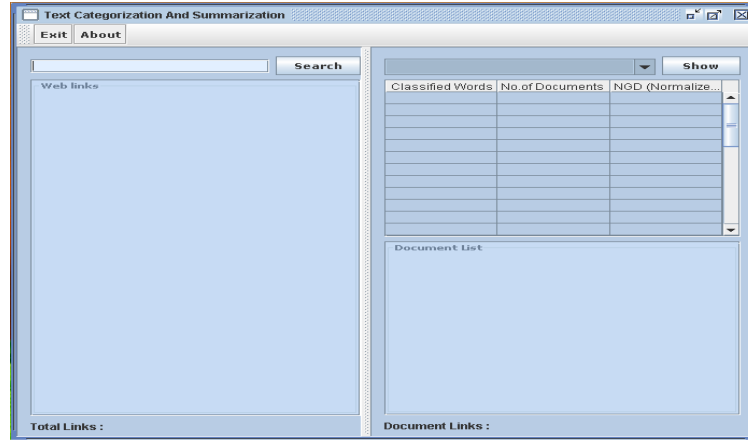

Figure 5.   Initial screen when the application starts

The first two "Text Categorization" and "Document List" stages of the model are implemented. Fig. 5 shows the snapshot of the initial screen when the application starts. On the left of the screen is a text box and search button where the user types a keyword and clicks on the search button to search the keyword on the web. Fig. 6 shows the result of searching the keyword "sports". And Fig. 7 shows the screen that displays the categories found related to the keyword, the number of documents found related to the keyword and the NGD (Normalized Google Distance) and it also displays the list of documents.

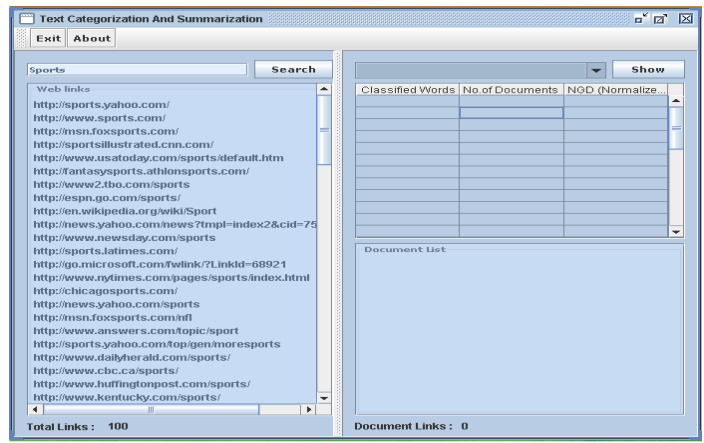

Figure 6.   Screen after user searches a keyword

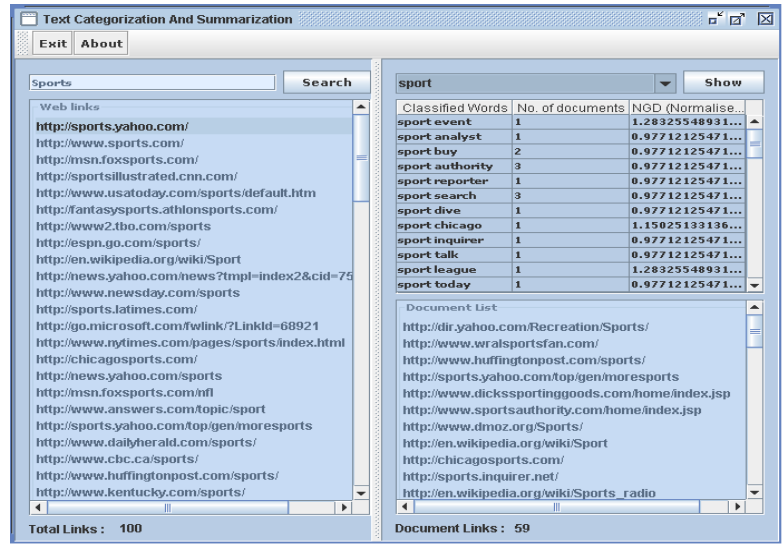

Figure 7. Screen after user clicks on show button

## VII. Conclusion

This model uses QDC algorithm for text categorization which makes it more powerful as QDC algorithm is evaluated against other clustering algorithms in [4]. By using text summarization after searching the document saves the user's time required for reading the complete document.

### Acknowledgment

Khushboo Thakkar thanks the Department of Computer Science and Engineering, G. H. Raisoni College of Engineering and Ms. Urmila Shrawankar for her special guidance.

**AUTHORS PROFILE**

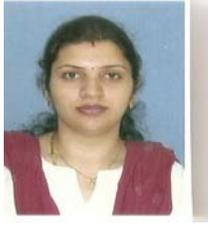

Khushboo Thakkar is a research student, pursuing M.Tech degree from G. H. Raisoni College of Engineering, Nagpur. And has done B.E. from Shri Ramdeobaba Kamla Nehru Engineering College, Nagpur.

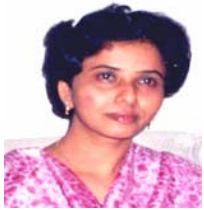

Ms.Urmila Shrawankar is Head of Department in G. H. Raisoni College of Engineering, Nagpur. Pursuing Ph.D. (CSE). Her area of specialization are Operating System and Human Computer Interaction.